\documentclass[12pt]{article}
\begin{document}

\def\ds{\displaystyle}
\def\beq{\begin{equation}}
\def\eeq{\end{equation}}
\def\bea{\begin{eqnarray}}
\def\eea{\end{eqnarray}}
\def\beeq{\begin{eqnarray}}
\def\eeeq{\end{eqnarray}}
\def\ve{\vert}
\def\vel{\left|}
\def\ver{\right|}
\def\nnb{\nonumber}
\def\ga{\left(}
\def\dr{\right)}
\def\aga{\left\{}
\def\adr{\right\}}
\def\lla{\left<}
\def\rra{\right>}
\def\rar{\rightarrow}
\def\nnb{\nonumber}
\def\la{\langle}
\def\ra{\rangle}
\def\ba{\begin{array}}
\def\ea{\end{array}}
\def\tr{\mbox{Tr}}
\def\ssp{{\Sigma^{*+}}}
\def\sso{{\Sigma^{*0}}}
\def\ssm{{\Sigma^{*-}}}
\def\xis0{{\Xi^{*0}}}
\def\xism{{\Xi^{*-}}}
\def\qs{\la \bar s s \ra}
\def\qu{\la \bar u u \ra}
\def\qd{\la \bar d d \ra}
\def\qq{\la \bar q q \ra}
\def\gGgG{\la g^2 G^2 \ra}
\def\q{\gamma_5 \not\!q}
\def\x{\gamma_5 \not\!x}
\def\g5{\gamma_5}
\def\sb{S_Q^{cf}}
\def\sd{S_d^{be}}
\def\su{S_u^{ad}}
\def\ss{S_s^{??}}
\def\sbp{{S}_Q^{'cf}}
\def\sdp{{S}_d^{'be}}
\def\sup{{S}_u^{'ad}}
\def\ssp{{S}_s^{'??}}
\def\sig{\sigma_{\mu \nu} \gamma_5 p^\mu q^\nu}
\def\fo{f_0(\frac{s_0}{M^2})}
\def\ffi{f_1(\frac{s_0}{M^2})}
\def\fii{f_2(\frac{s_0}{M^2})}
\def\O{{\cal O}}
\def\sl{{\Sigma^0 \Lambda}}
\def\es{\!\!\! &=& \!\!\!}
\def\ap{\!\!\! &\approx& \!\!\!}
\def\ar{&+& \!\!\!}
\def\ek{&-& \!\!\!}
\def\kek{\!\!\!&-& \!\!\!}
\def\cp{&\times& \!\!\!}
\def\se{\!\!\! &\simeq& \!\!\!}
\def\eqv{&\equiv& \!\!\!}
\def\kpm{&\pm& \!\!\!}
\def\kmp{&\mp& \!\!\!}


\def\simlt{\stackrel{<}{{}_\sim}}
\def\simgt{\stackrel{>}{{}_\sim}}


\title{
         {\Large
                 {\bf
Double--lepton polarization asymmetries in
$\Lambda_b \rar \Lambda \ell^+ \ell^-$ decay in universal 
extra dimension model
                 }
         }
      }

\author{\vspace{1cm}\\
{
\small T. M. Aliev$^{(1)}$ \thanks
{e-mail: taliev@metu.edu.tr}~\footnote{permanent address: Institute
of Physics, Baku, Azerbaijan}\,\,,
M. Savc{\i}$^{(1)}$ \thanks
{e-mail: savci@metu.edu.tr}\,\,,
B. B. \c{S}irvanl{\i}$^{(2)}$ \thanks
{e-mail: bbelma@gazi.edu.tr} 
} \\
{\small $(1)$ Physics Department, Middle East Technical University,
06531 Ankara, Turkey}\\
\small $(2)$ Physics Department, Gazi University,
06500 Teknik Okullar, Ankara, Turkey }

\date{}

\begin{titlepage}
\maketitle
\thispagestyle{empty}

\begin{abstract}
Double--lepton polarization asymmetries in $\Lambda_b \rar \Lambda \ell^+ 
\ell^-$ decay are calculated in universal extra dimension (UED) model. It is
obtained that numerous double--lepton polarization asymmetries are very
sensitive to the UED model and therefore can be very useful tool for
establishing new physics predicted by the UED model.
\end{abstract}

~~~PACS numbers: 12.60.--i, 13.30.--a, 14.20.Mr
\end{titlepage}

\section{Introduction}

Despite the impressive success of the standard model (SM) in describing all
existing experimental data, it is commonly believed that SM is the low
energy limit of a more fundamental theory. There are two different ways in
looking the evidence for new physics beyond the SM:
\begin{itemize}
\item direct production of new particles at high energy colliders like LHC;

\item signals of new interactions and particles can be obtained indirectly
through the analysis of rare decays.
\end{itemize}

Rare B meson decays induced by the $b \rar s(d)$ transitions play a special
role, since they are forbidden at at tree level in the SM and appear only at
quantum (one--loop) level. Moreover, these decays are the most promising ones for
establishing new physics. New physics in these decays can appear either
through the differences in the Wilson coefficients from the ones existing in
the SM or through the new operator structures in the effective Hamiltonian
which are absent in the SM.  

Among all decay channels of B mesons, semileptonic ones receive a special
interest. These decays are theoretically, more or less,
clean and they have relatively larger branching ratio. These decays contain 
many physically measurable quantities, like forward--backward asymmetry 
${\cal A}_{FB}$, lepton
polarization asymmetries, etc., which are very useful and serve as a testing
ground for the SM and looking for new physics beyond the SM \cite{R7701}.
From experimental side, BELLE \cite{R7702,R7703} and BaBar
\cite{R7704,R7705} collaborations provide recent measurements of the
branching ratios of the semileptonic decays due to the $b \rar s \ell^+
\ell^-$ transitions, which can be summarized as:
\bea
{\cal B}(B \rar K^\ast \ell^+ \ell^-) \es \left\{ \begin{array}{lc}
\left( 16.5^{+2.3}_{-2.2} \pm 0.9 \pm 0.4\right) \times
10^{-7}& \cite{R7702}~,\\ \\
\left( 7.8^{+1.9}_{-1.7} \pm 1.2\right) \times
10^{-7}& \cite{R7704}~,\end{array} \right.
\nnb \\ \nnb \\
{\cal B}(B \rar K \ell^+ \ell^-) \es \left\{ \begin{array}{lc}
\left( 5.5^{+0.75}_{-0.70} \pm 0.27 \pm 0.02\right) \times
10^{-7}& \cite{R7702}~,\\ \\
\left( 3.4  \pm 0.7 \pm 0.3\right) \times
10^{-7}& \cite{R7704}~.\end{array} \right.
\nnb \\ \nnb \\
{\cal B}(B \rar X_s \ell^+ \ell^-) \es \left\{ \begin{array}{lc}
\left( 4.11 \pm 0.83^{+0.85}_{-0.81}\right) \times
10^{-6}& \cite{R7703}~,\\ \\
\left( 5.6  \pm 1.5 \pm 0.6 \pm 1.1 \right) \times
10^{-6}& \cite{R7705}~.\end{array} \right. \nnb
\eea

Another exclusive decay which is described at inclusive level by the $b \rar
s \ell^+ \ell^-$ transition is the baryonic $\Lambda_b \rar \Lambda \ell^+
\ell^-$ decay. Unlike mesonic decays, the baryonic decays could maintain the
helicity structure of the effective Hamiltonian for the $b \rar s$
transition \cite{R7706}. Radiative and semileptonic decays of $\Lambda_b$
such as $\Lambda_b \rar \Lambda \gamma$, $\Lambda_b \rar \Lambda_c \ell
\bar{\nu}_\ell$, $\Lambda_b \rar \Lambda \ell^+ \ell^-$ $(\ell = e,\mu,\tau)$
and $\Lambda_b \rar \Lambda \nu \bar{\nu}$ have been extensively studied in
the literature \cite{R7707} (see also \cite{R7701} and references therein). 
More about heavy baryons, including the experimental prospects, can be found in
\cite{R7708,R7709}. 

It is noted in \cite{R7710} that some of the single lepton polarization
asymmetries might be too small to be observed and therefore might not provide
sufficient number of observables for checking the structure of the effective
Hamiltonian. In order to obtain more observables, London et. al., proposed
to take polarizations of both leptons into account \cite{R7710} which are
simultaneously measurable. Along these lines maximum number of independent
polarization observables are constructed in \cite{R7710}.    

Among the various models of physics beyond the SM, extra dimensions
attract special interest, because they include gravity in addition other
interactions, giving hints on the hierarchy problem and a connection with
string theory. The model of Appelquist, Cheng and Dobrescu (ACD)
\cite{R7711} with one universal extra dimension (UED), where all the SM
particles can propagate in the extra dimension, are very attractive (see
also \cite{R7712}). Compactification of the extra dimension leads to 
Kaluza--Klein model in the four--dimension. In this model the only 
additional free parameter with respect to the SM is $1/R$, i.e., inverse 
of the compactification radius.

The restrictions imposed on UED are examined in the current accelerators,
for example, Tevatron experiments put the bound about $1/R\ge 300~GeV$.
Analysis of the anomalous magnetic moment \cite{R7713}, and $Z \rar \bar{b}
b$ vertex \cite{R7714} also lead to the bound $1/R\ge 300~GeV$.

Possible manifestation of UED models in the $K_L$--$K_S$ mass difference,
parameter $\varepsilon_K$, $B$--$\bar{B}_0$ mixing, $\Delta M_{d,s}$ mass
difference, and rare decays $K^+ \rar \pi \bar{\nu} \nu$ , $K_L \rar \pi^0
\bar{\nu} \nu$, $K_L \rar \mu^+ \mu^-$, $B \rar X_{s,d} \bar{\nu} \nu$, 
$B_{s,d} \rar \mu^+ \mu^-$, $B \rar X_s \gamma$, $B \rar X_s ~gluon$,
$B \rar X_s \mu^+ \mu^-$ and $\varepsilon^\prime/\varepsilon$ are
comprehensively investigated in \cite{R7715} and \cite{R7716}. Exclusive 
$B \rar K^\ast \ell^+ \ell^-$, $B \rar K^\ast \bar{\nu} \nu$ and $B \rar
K^\ast \gamma$ decays are studied in the framework of the UED scenario in
\cite{R7717}, and $\Lambda_b \rar \Lambda \ell^+ \ell^-$ in UED in
\cite{R7718}.  

In the present work we study the double--lepton polarization asymmetries for
the $\Lambda_b \rar \Lambda \ell^+ \ell^-$ decay in the UED model. 
The plan of the paper is as follows. In section
2 we briefly discuss the main ingredients of ACD model and
calculate all possible double--lepton polarization asymmetries
for the rare $\Lambda_b \rar \Lambda \ell^+ \ell^-$ decay.
Section 3 is devoted to the numerical analysis and conclusions.

\section{$\Lambda_b \rar \Lambda \ell^+ \ell^-$ decay in 
ADC model}

Let us remind the interested reader about the main ingredients of
the simplest ACD model, which is the minimal extension of the SM in $4+1$
dimensions. The five--dimensional ACD model with a single UED uses orbifold
compactification, namely, the fifth dimension $y$ that is compactified in a circle of
radius $R$, with points $y=0$ and $y=\pi R$ that are fixed points of the
orbifolds. Generalization of the SM is realized by the propagating fermions, 
gauge bosons and the Higgs fields in all five dimensions. The Lagrangian 
in ACD can be written as
\bea
{\cal L} = \int d^4x dy \Big\{ {\cal L}_A + {\cal L}_H + {\cal L}_F +
{\cal L}_Y \Big\}~, \nnb
\eea
where
\bea
{\cal L}_A \es - \frac{1}{4} W^{MNa} W_{MN}^a - 
\frac{1}{4} B^{MN} B_{MN}~, \nnb \\   
{\cal L}_H \es \Big( {\cal D}^M \phi \Big)^\dagger {\cal D}_M \phi 
- V(\phi)~, \nnb \\
{\cal L}_F \es \bar{ {\cal Q} } \Big( i \Gamma^M {\cal D}_M \Big) {\cal Q} +
\bar{u} \Big( i \Gamma^M {\cal D}_M \Big) u +
\bar{ {\cal D} } \Big( i \Gamma^M {\cal D}_M \Big) {\cal D}~, \nnb \\
{\cal L}_Y \es - \bar{ {\cal Q} } \widetilde{Y}_u \phi^c u - 
\bar{ {\cal Q} } \widetilde{Y}_d \phi {\cal D} + \mbox{\rm h.c.}~.\nnb
\eea
Here $M$ and $N$ running over 0,1,2,3,5 are the five--dimensional Lorentz
indices, $W_{MN}^a= {\partial}_M W_N^a - {\partial}_N W_M^a + \tilde{g}
\varepsilon^{abc} W_M^b W_N^c$ are the field strength tensor for the
$SU(2)_L$ electroweak gauge group, $B_{MN}={\partial}_M B_N -
{\partial}_N B_M$ are that of the $U(1)$ group, and all fields depend both on
$x$ and $y$. The covariant derivative is defined as ${\cal D}_M =
{\partial}_M - i \tilde{g} W_M^a T^a - i\tilde{g}^\prime B_M Y$, where 
$\tilde{g}$ and $\tilde{g}^\prime$ are the five--dimensional gauge couplings
for the $SU(2)_L$ and $U(1)$ groups. The five--dimensional $\Gamma_M$
matrices are defined as $\Gamma^\mu=\gamma^\mu~,~\mu=0,1,2,3$ and $\Gamma^5
= i \gamma^5$.

In the case o a single extra dimension with coordinate $x_5=y$ compactified
on a circle of radius $R$, a field $F(x,y)$ would be periodic function of
$y$, hence can be written as
\bea
F(x,y)=\sum_{n=-\infty}^{+\infty} F_n(x) e^{iny/R}~.\nnb
\eea

The Fourier expansion of the fields are
\bea
B_\mu(x,y) \es \frac{1}{\sqrt{2 \pi R}} B_\mu^{(0)} + 
\frac{1}{\sqrt{\pi R}} \sum_{n=1}^{\infty} B_\mu^{(0)} (x) 
\cos \left(\frac{ny}{R} \right)~, \nnb \\
B_5(x,y) \es \frac{1}{\sqrt{\pi R}} \sum_{n=1}^{\infty} B_5^{(n)} 
\sin \left(\frac{ny}{R} \right)~, \nnb \\
{\cal Q}(x,y) \es \frac{1}{\sqrt{2 \pi R}} {\cal Q}_L^{(0)} +
\frac{1}{\sqrt{\pi R}} \sum_{n=1}^{\infty} \left[
{\cal Q}_L^{(n)} \cos \left(\frac{ny}{R} \right) +
{\cal Q}_R^{(n)} \sin \left(\frac{ny}{R} \right) \right]~,\nnb \\
U({\cal  D})(x,y) \es \frac{1}{\sqrt{2 \pi R}} U_R^{(0)} +
\frac{1}{\sqrt{\pi R}} \sum_{n=1}^{\infty} \left[
U_R^{(n)} \cos \left(\frac{ny}{R} \right) + 
U_L^{(n)} \sin \left(\frac{ny}{R} \right) \right]~.\nnb
\eea
Under parity transformation $P_5:y \rar -y$ fields having a correspondent in
the four--dimensional SM should be even, so that their zero--mode in the
KK can be interpreted as the ordinary SM field, and all remaining new fields
should be odd.

In ACD model the KK parity is conserved. This
conservation implies that there is no tree level
diagrams with exchange of KK modes in low energy processes (at the scale 
$\mu \ll 1/R$) and single KK excitation cannot be produced, i.e., they
appear only in pairs. Lastly, in the ACD model there are three
additional physical scalar modes $a_n^{(0)}$ and $a_n^\pm$.
The zero--mode is either right--handed or left--handed.

Lagrangian of the ACD model can be obtained by integrating over $x_5=y$
\bea
{\cal L}_4 (x) = \int_0^{2\pi R} {\cal L}_5 (x,y) dy~.\nnb
\eea
Note that the zero--mode remains massless unless we apply the Higgs
mechanism. All fields in the four--dimensional Lagrangian receive the KK
mass $n/R$ on account of the derivative operator ${\partial}_5$ acting on
them.The relevant Feynman rules are derived in \cite{R7715} and for more
details about the ACD model we refer the interested reader to \cite{R7716}
and \cite{R7717}.

After this introduction, let us start discussing the main
problem, namely, double--lepton polarization asymmetries 
for the $\Lambda_b \rar \Lambda \ell^+ \ell^-$ decay.

At quark level, $\Lambda_b \rar \Lambda \ell^+ \ell^-$ decay is described by
$b \rar s \ell^+ \ell^-$ transition. Effective Hamiltonian governing this
transition in the SM with $\Delta B=-1$, $\Delta S = 1$ is described in
terms of a set of local operators
\bea
\label{e7701}
{\cal H} = \frac{4 G_F}{\sqrt{2}} V_{tb} V^\ast_{ts} \sum_1^{10} C_i(\mu)
{\cal O}_i(\mu)~,
\eea
where $G_F$ is the Fermi constant, $V_{ij}$ are the elements of the
Cabibbo--Kobayashi--Maskawa (CKM) matrix. Explicit forms of the operators,
which are written in terms of quark and gluon fields can be found in
\cite{R7719}.

The Wilson coefficients in (\ref{e7701}) have been computed at NNLO in the
SM in \cite{R7719}. At NLO the coefficients are calculated for the ACD model
including the effects of KK modes, in \cite{R7715} and \cite{R7716}, which
we have used in our calculations. It should be noted here that, there does not
appear any new operator in the ACD model, and therefore, the effect of new
particles leads to modification of the Wilson coefficients existing in the SM, if we
neglect the contributions of the scalar fields, which are indeed very small. 

At $\mu={\cal O}(m_W)$ level, only $C_2^{(0)}$, $C_7^{(0)(m_W)}$, $C_8^{(0)(m_W)}$,
$C_9^{(0)(m_W)}$ and $C_{10}^{(0)(m_W)}$ are different from zero, and the
remaining coefficients are all zero.

In the SM, at quark level, $\Lambda_b \rar \Lambda \ell^+ \ell^-$ decay is 
described with the help of the operators $C_7^{eff}$, $C_9$ and $C_{10}$ as
follows:
\bea
\label{e7702}
{\cal M} \es \frac{G_F}{4 \sqrt{2}} V_{tb} V^\ast_{ts} \Big\{
C_7^{eff} \bar{s} i \sigma_{\mu\nu} (1+ \gamma_5) q^\nu b \bar{\ell}\gamma^\mu \ell
+ C_9 \bar{s} \gamma_\mu (1-\gamma_5) b \bar{\ell} \gamma^\mu \ell \nnb \\
\ar C_{10} \bar{s} \gamma_\mu (1-\gamma_5) b \bar{\ell} \gamma^\mu 
\gamma_5 \ell \Big\}~.
\eea   

As has already been noted,
$C_7^{eff}$, $C_9$ and $C_{10}$ are calculated in the SM
in \cite{R7719} (see also \cite{R7720} and \cite{R7721}).

Contributions coming from UED model to these Wilson coefficients are
calculated in \cite{R7715} and \cite{R7716}, which can be written as

\bea 
\label{e7703}
C_7^{(0)}(\mu_W) \es -\frac{1}{2} D^\prime(x_t,1/R)~,\nnb \\
C_9(\mu) \es P_0^{NDR}+{Y(x_t,1/R) \over \sin^2 \theta_W} -4
Z(x_t,1/R)+P_E E(x_t,1/R)~, \nnb \\
C_{10} \es - \frac{Y(x_t,1/R)}{\sin^2 \theta_W}~.
\eea
where $P_0^{NDR}=2.60 \pm 0.25$ and the superscript $(0)$ referring to 
leading log approximation. Explicit expressions of the functions
$D^\prime(x_t,1/R)$, $Y(x_t,1/R)$ and $Z(x_t,1/R)$ can be found in
\cite{R7715,R7716,R7717}.  

With these coefficients and the operators in (\ref{e7701}) the inclusive 
$b \to s \ell^+ \ell^-$ transitions are studied in \cite{R7715,R7716}.  

The amplitude of the exclusive $\Lambda_b \rar \Lambda\ell^+ \ell^-$ decay
is obtained by calculating the matrix element of the effective Hamiltonian for the 
$b \rar s \ell^+ \ell^-$ transition between initial and final
baryon states $\lla \Lambda \vel {\cal H}_{eff} \ver \Lambda_b \rra$.
It follows from Eq. (\ref{e7702}) that the matrix elements
\bea
\label{e7704}
&&\lla \Lambda \vel \bar s \gamma_\mu (1 - \gamma_5) b \ver \Lambda_b
\rra~,\nnb \\
&&\lla \Lambda \vel \bar s \sigma_{\mu\nu} (1 + \gamma_5) b \ver \Lambda_b
\rra~,
\eea
are needed in order to calculate
the $\Lambda_b \rar \Lambda\ell^+ \ell^-$ decay amplitude.

These matrix elements parametrized in terms of the form factors are 
as follows (see \cite{R7722,R7723})
\bea
\label{e7705}
\lla \Lambda \vel \bar s \gamma_\mu b \ver \Lambda_b \rra  
\es \bar u_\Lambda \Big[ f_1 \gamma_\mu + i f_2 \sigma_{\mu\nu} q^\nu + f_3  
q_\mu \Big] u_{\Lambda_b}~,\\
\label{e7706}
\lla \Lambda \vel \bar s \gamma_\mu \gamma_5 b \ver \Lambda_b \rra
\es \bar u_\Lambda \Big[ g_1 \gamma_\mu \gamma_5 + i g_2 \sigma_{\mu\nu}
\gamma_5 q^\nu + g_3 q_\mu \gamma_5\Big] u_{\Lambda_b}~,
\eea
where $q= p_{\Lambda_b} - p_\Lambda$. 

The form factors of the magnetic dipole operators are defined as 
\bea
\label{e7707}
\lla \Lambda \vel \bar s i \sigma_{\mu\nu} q^\nu  b \ver \Lambda_b \rra
\es \bar u_\Lambda \Big[ f_1^T \gamma_\mu + i f_2^T \sigma_{\mu\nu} q^\nu
+ f_3^T q_\mu \Big] u_{\Lambda_b}~,\nnb \\
\lla \Lambda \vel \bar s i \sigma_{\mu\nu}\gamma_5  q^\nu  b \ver \Lambda_b \rra
\es \bar u_\Lambda \Big[ g_1^T \gamma_\mu \gamma_5 + i g_2^T \sigma_{\mu\nu}
\gamma_5 q^\nu + g_3^T q_\mu \gamma_5\Big] u_{\Lambda_b}~.
\eea

Using the identity 
\bea
\sigma_{\mu\nu}\gamma_5 = - \frac{i}{2} \epsilon_{\mu\nu\alpha\beta}
\sigma^{\alpha\beta}~,\nnb
\eea
the following relations between the form factors are obtained:
\bea
\label{e7708}
f_1^T \es - \frac{q^2}{m_{\Lambda_b} - m_\Lambda} f_3^T~,\nnb \\
g_1^T \es \frac{q^2}{m_{\Lambda_b} + m_\Lambda} g_3^T~.
\eea 

Using these definitions of the form factors, for the matrix element
of the $\Lambda_b \rar \Lambda\ell^+ \ell^-$ we get
\bea
\label{e7709}
{\cal M} \es \frac{G \alpha}{4 \sqrt{2}\pi} V_{tb}V_{ts}^\ast \frac{1}{2} \Bigg\{
\bar{\ell} \gamma_\mu (1-\gamma_5) \ell \, 
\bar{u}_\Lambda \Big[ (A_1 - D_1) \gamma_\mu (1+\gamma_5) +
(B_1 - E_1) \gamma_\mu (1-\gamma_5) \nnb \\
\ar i \sigma_{\mu\nu} q^\nu \Big( (A_2 - D_2) (1+\gamma_5) +
(B_2 - E_2) (1-\gamma_5) \Big) \nnb \\
\ar q_\mu \Big( (A_3 - D_3) (1+\gamma_5) + (B_3 - E_3) (1-\gamma_5)
\Big) \Big] u_{\Lambda_b} \nnb \\
\ar \bar{\ell} \gamma_\mu (1+\gamma_5) \ell \, 
\bar{u}_\Lambda \Big[ (A_1 + D_1) \gamma_\mu (1+\gamma_5) +
(B_1 + E_1) \gamma_\mu (1-\gamma_5) \nnb \\
\ar i \sigma_{\mu\nu} q^\nu \Big( (A_2 + D_2) (1+\gamma_5) +
(B_2 + E_2) (1-\gamma_5) \Big) \nnb \\
\ar q_\mu \Big( (A_3 + D_3) (1+\gamma_5) + (B_3 + E_3) (1-\gamma_5) \Big)
\Big] u_{\Lambda_b} \Bigg\}~,
\eea
where
\bea
\label{e7710}
A_1 \es \frac{1}{q^2}\ga f_1^T-g_1^T \dr (-2 m_s C_7) + \frac{1}{q^2}\ga
f_1^T+g_1^T \dr (-2 m_b C_7) + \ga f_1-g_1 \dr C_9^{eff}~,\nnb \\
A_2 \es A_1 \ga 1 \rar 2 \dr ~,\nnb \\
A_3 \es A_1 \ga 1 \rar 3 \dr ~,\nnb \\
B_1 \es A_1 \ga g_1 \rar - g_1;~g_1^T \rar - g_1^T \dr ~,\nnb \\
B_2 \es B_1 \ga 1 \rar 2 \dr ~,\nnb \\
B_3 \es B_1 \ga 1 \rar 3 \dr ~,\nnb \\
D_1 \es C_{10} \ga f_1-g_1 \dr~,\nnb \\
D_2 \es D_1 \ga 1 \rar 2 \dr ~, \\
D_3 \es D_1 \ga 1 \rar 3 \dr ~,\nnb \\
E_1 \es D_1 \ga g_1 \rar - g_1 \dr ~,\nnb \\
E_2 \es E_1 \ga 1 \rar 2 \dr ~,\nnb \\
E_3 \es E_1 \ga 1 \rar 3 \dr ~.\nnb
\eea

From these expressions it follows
that $\Lambda_b \rar\Lambda \ell^+\ell^-$ decay is described in terms of  
many form factors. It is shown in \cite{R7706} (see also \cite{R7723}) 
that Heavy Quark Effective Theory reduces
the number of independent form factors to two ($F_1$ and
$F_2$) irrelevant of the Dirac structure
of the corresponding operators, i.e., 
\bea
\label{e7711}
\lla \Lambda(p_\Lambda) \vel \bar s \Gamma b \ver \Lambda(p_{\Lambda_b})
\rra = \bar u_\Lambda \Big[F_1(q^2) + \not\!v F_2(q^2)\Big] \Gamma
u_{\Lambda_b}~,
\eea
where $\Gamma$ is an arbitrary Dirac structure and
$v^\mu=p_{\Lambda_b}^\mu/m_{\Lambda_b}$ is the four--velocity of
$\Lambda_b$. Comparing the general form of the form factors given in Eqs.
(\ref{e7705})--(\ref{e7707}) with the ones given in (\ref{e7711}), one can
easily obtain the following relations among them \cite{R7722,R7723},
\bea
\label{e7712}
g_1 \es f_1 = f_2^T= g_2^T = F_1 + \sqrt{\hat{r}_\Lambda} F_2~, \nnb \\
g_2 \es f_2 = g_3 = f_3 = \frac{F_2}{m_{\Lambda_b}}~,\nnb \\
g_1^T \es f_1^T = \frac{F_2}{m_{\Lambda_b}} q^2~,\nnb \\
g_3^T \es \frac{F_2}{m_{\Lambda_b}} \ga m_{\Lambda_b} + m_\Lambda \dr~,\nnb \\
f_3^T \es - \frac{F_2}{m_{\Lambda_b}} \ga m_{\Lambda_b} - m_\Lambda \dr~,
\eea
where $\hat{r}_\Lambda=m_\Lambda^2/m_{\Lambda_b}^2$.

As we have already noted, our purpose is the calculation of double--lepton
polarizations in UED model.

For calculation of the double lepton polarization asymmetries, 
the following orthogonal unit vectors
$s_i^{\pm\mu}$ in the rest frame of $\ell^\pm$ 
($i=L,T$ or $N$, stand for longitudinal, transversal or
normal polarizations, respectively, are chosen as:

\bea
\label{e7713}   
s^{-\mu}_L \es \ga 0,\vec{e}_L^{\,-}\dr =
\ga 0,\frac{\vec{p}_-}{\vel\vec{p}_- \ver}\dr~, \nnb \\
s^{-\mu}_N \es \ga 0,\vec{e}_N^{\,-}\dr = \ga 0,\frac{\vec{p}_\Lambda\times
\vec{p}_-}{\vel \vec{p}_\Lambda\times \vec{p}_- \ver}\dr~, \nnb \\
s^{-\mu}_T \es \ga 0,\vec{e}_T^{\,-}\dr = \ga 0,\vec{e}_N^{\,-}
\times \vec{e}_L^{\,-} \dr~, \nnb \\
s^{+\mu}_L \es \ga 0,\vec{e}_L^{\,+}\dr =
\ga 0,\frac{\vec{p}_+}{\vel\vec{p}_+ \ver}\dr~, \nnb \\
s^{+\mu}_N \es \ga 0,\vec{e}_N^{\,+}\dr = \ga 0,\frac{\vec{p}_\Lambda\times
\vec{p}_+}{\vel \vec{p}_\Lambda\times \vec{p}_+ \ver}\dr~, \nnb \\
s^{+\mu}_T \es \ga 0,\vec{e}_T^{\,+}\dr = \ga 0,\vec{e}_N^{\,+}
\times \vec{e}_L^{\,+}\dr~,
\eea
where $\vec{p}_\mp$ and $\vec{p}_\Lambda$ are the three--momenta of the
leptons $\ell^\mp$ and $\Lambda$ baryon in the
center of mass frame (CM) of $\ell^- \,\ell^+$ system, respectively.
Transformation of unit vectors from the rest frame of the leptons to CM
frame of leptons can be done by the Lorentz boost. Boosting of the
longitudinal unit vectors $s_L^{\pm\mu}$ yields
\bea
\label{e7714}
\ga s^{\mp\mu}_L \dr_{CM} \es \ga \frac{\vel\vec{p}_\mp \ver}{m_\ell}~,
\frac{E_\ell \vec{p}_\mp}{m_\ell \vel\vec{p}_\mp \ver}\dr~,
\eea
where $\vec{p}_+ = - \vec{p}_-$, $E_\ell$ and $m_\ell$ are the energy and mass
of leptons in the CM frame, respectively.
The remaining two unit vectors $s_N^{\pm\mu}$, $s_T^{\pm\mu}$ are unchanged
under Lorentz boost.

The double--polarization asymmetries are defined in the following way 
\cite{R7710}:

\bea                                                                  
\label{e7715}
P_{ij}(q^2) \es
\frac{ 
\Big( \ds \frac{d\Gamma(\vec{s}^{\,-}_i,\vec{s}^{\,+}_j)}{dq^2} -
      \ds \frac{d\Gamma(-\vec{s}^{\,-}_i,\vec{s}^{\,+}_j)}{dq^2} \Big) -
\Big( \ds \frac{d\Gamma(\vec{s}^{\,-}_i,-\vec{s}^{\,+}_j)}{dq^2} -      
      \ds \frac{d\Gamma(-\vec{s}^{\,-}_i,-\vec{s}^{\,+}_j)}{dq^2} \Big) 
     }
     {    
\Big( \ds \frac{d\Gamma(\vec{s}^{\,-}_i,\vec{s}^{\,+}_j)}{dq^2} +      
      \ds \frac{d\Gamma(-\vec{s}^{\,-}_i,\vec{s}^{\,+}_j)}{dq^2} \Big) +
\Big( \ds \frac{d\Gamma(\vec{s}^{\,-}_i,-\vec{s}^{\,+}_j)}{dq^2} +      
      \ds \frac{d\Gamma(-\vec{s}^{\,-}_i,-\vec{s}^{\,+}_j)}{dq^2} \Big)
     }~,
\eea
where, the first subindex $i$ represents lepton and the second one
antilepton. Using this definition of $P_{ij}$, nine double--lepton
polarization asymmetries are calculated. Their expressions are

\bea
\label{e7716}
P_{LL} \es \frac{16 m_{\Lambda_b}^4}{3 \Delta}
\mbox{\rm Re} \Bigg\{
- 6 m_{\Lambda_b} \sqrt{\hat{r}_\Lambda}
(1-\hat{r}_\Lambda+\hat{s})
\Big[ \hat{s} (1+v^2) (A_1 A_2^\ast + B_1 B_2^\ast)  - 
4 \hat{m}_\ell^2 (D_1 D_3^\ast + E_1 E_3^\ast) \Big] \nnb \\
\ar 6 m_{\Lambda_b} (1-\hat{r}_\Lambda-\hat{s})
\Big[ \hat{s} (1+v^2) (A_1 B_2^\ast + A_2 B_1^\ast) + 
4 \hat{m}_\ell^2 (D_1 E_3^\ast + D_3 E_1^\ast) \Big] \nnb \\
\ar 12 \sqrt{\hat{r}_\Lambda} \hat{s} (1+v^2) 
\Big( A_1 B_1^\ast + D_1 E_1^\ast +
m_{\Lambda_b}^2 \hat{s} A_2 B_2^\ast \Big) \nnb \\
\ar 12 m_{\Lambda_b}^2 \hat{m}_\ell^2 \hat{s} (1+\hat{r}_\Lambda-\hat{s})
\ga \vel D_3 \ver^2 + \vel E_3^\ast \ver^2 \dr \nnb \\
\ek (1+v^2)
\Big[ 1+\hat{r}_\Lambda^2 - 
\hat{r}_\Lambda (2-\hat{s}) +\hat{s} (1-2 \hat{s}) \Big]
\Big(\vel A_1 \ver^2 + \vel B_1 \ver^2 \Big) \nnb \\
\ek \Big[
(5 v^2 - 3) (1-\hat{r}_\Lambda)^2 +   
4 \hat{m}_\ell^2 (1+\hat{r}_\Lambda) +
2 \hat{s} (1+8 \hat{m}_\ell^2 + \hat{r}_\Lambda)
- 4 \hat{s}^2 \Big] \Big( \vel D_1 \ver^2 + \vel E_1 \ver^2 \Big) \nnb \\
\ek m_{\Lambda_b}^2 (1+v^2) \hat{s}
\Big[2 + 2 \hat{r}_\Lambda^2 -\hat{s}(1 +\hat{s}) -
\hat{r}_\Lambda (4 + \hat{s})\Big] \big(
\vel A_2 \ver^2 + \vel B_2 \ver^2 \Big) \nnb \\
\ek 2 m_{\Lambda_b}^2 \hat{s} v^2 \Big[
2 (1 + \hat{r}_\Lambda^2) - \hat{s} (1+\hat{s}) - 
\hat{r}_\Lambda (4+\hat{s})\Big] \Big(
\vel D_2 \ver^2 + \vel E_2 \ver^2 \Big) \nnb \\
\ar 12 m_{\Lambda_b} \hat{s} (1-\hat{r}_\Lambda-\hat{s}) v^2
\Big( D_1 E_2^\ast + D_2 E_1^\ast \Big) \nnb \\
\ek 12 m_{\Lambda_b} \sqrt{\hat{r}_\Lambda} \hat{s}
(1-\hat{r}_\Lambda+\hat{s}) v^2
\Big( D_1 D_2^\ast + E_1 E_2^\ast \Big) \nnb \\
\ar 24 m_{\Lambda_b}^2 \sqrt{\hat{r}_\Lambda} \hat{s}
\Big( \hat{s} v^2 D_2 E_2^\ast + 
2 \hat{m}_\ell^2 D_3 E_3^\ast \Big)
\Bigg\}~, \\ \nnb \\ \nnb \\
\label{e7717}
P_{LN} \es \frac{16 \pi m_{\Lambda_b}^4 \hat{m}_\ell \sqrt{\lambda}}{\Delta \sqrt{\hat{s}}} 
\mbox{\rm Im} \Bigg\{
(1-\hat{r}_\Lambda) 
(A_1^\ast D_1 + B_1^\ast E_1)
+ m_{\Lambda_b}  
 \hat{s} (A_1^\ast E_3 - A_2^\ast E_1 + B_1^\ast D_3
-B_2^\ast D_1) \nnb \\
\ar m_{\Lambda_b}
 \sqrt{\hat{r}_\Lambda} \hat{s}
(A_1^\ast D_3 + A_2^\ast D_1 +B_1^\ast E_3 + B_2^\ast E_1)
- m_{\Lambda_b}^2 \hat{s}^2
\Big( B_2^\ast E_3 + A_2^\ast D_3 \Big)
\Bigg\}~, \\ \nnb \\ \nnb \\
\label{e7718}
P_{NL} \es - \frac{16 \pi m_{\Lambda_b}^4 \hat{m}_\ell \sqrt{\lambda}}{\Delta \sqrt{\hat{s}}} 
\mbox{\rm Im} \Bigg\{
(1-\hat{r}_\Lambda) 
(A_1^\ast D_1 + B_1^\ast E_1)
+ m_{\Lambda_b}   
 \hat{s} (A_1^\ast E_3 - A_2^\ast E_1 + B_1^\ast D_3
-B_2^\ast D_1) \nnb \\
\ek m_{\Lambda_b}
 \sqrt{\hat{r}_\Lambda} \hat{s}
(A_1^\ast D_3 + A_2^\ast D_1 +B_1^\ast E_3 + B_2^\ast E_1)
- m_{\Lambda_b}^2 \hat{s}^2
\Big( B_2^\ast E_3 + A_2^\ast D_3 \Big)
\Bigg\}~, \\ \nnb \\ \nnb \\
\label{e7719}
P_{LT} \es \frac{16 \pi m_{\Lambda_b}^4 \hat{m}_\ell \sqrt{\lambda} v}{\Delta \sqrt{\hat{s}}} 
\mbox{\rm Re} \Bigg\{
(1-\hat{r}_\Lambda) \Big( \vel D_1 \ver^2 + \vel E_1 \ver^2
\Big)
- \hat{s} \Big(A_1 D_1^\ast - B_1 E_1^\ast \Big) \nnb \\
\ek m_{\Lambda_b}
\hat{s} \Big[ B_1 D_2^\ast + (A_2 + D_2 -D_3) E_1^\ast
-  A_1 E_2^\ast
-(B_2-E_2+E_3) D_1^\ast \Big] \nnb \\
\ar m_{\Lambda_b}
 \sqrt{\hat{r}_\Lambda} \hat{s}
\Big[ A_1 D_2^\ast + (A_2 + D_2 +D_3) D_1^\ast - B_1 E_2^\ast -
(B_2 - E_2 - E_3) E_1^\ast \Big] \nnb \\ 
\ar m_{\Lambda_b}^2 
\hat{s} (1-\hat{r}_\Lambda)
(A_2 D_2^\ast - B_2 E_2^\ast)
- m_{\Lambda_b}^2             
\hat{s}^2
(D_2 D_3^\ast + E_2 E_3^\ast )
\Bigg\}~, \\ \nnb \\ \nnb \\
\label{e7720}
P_{TL} \es \frac{16 \pi m_{\Lambda_b}^4 \hat{m}_\ell \sqrt{\lambda} v}{\Delta \sqrt{\hat{s}}} 
\mbox{\rm Re} \Bigg\{
(1-\hat{r}_\Lambda) \Big( \vel D_1 \ver^2 + \vel E_1 \ver^2
\Big)
+ \hat{s} \Big(A_1 D_1^\ast - B_1 E_1^\ast \Big) \nnb \\
\ar m_{\Lambda_b} \hat{s}
\Big[ B_1 D_2^\ast + (A_2 - D_2 + D_3) E_1^\ast
-  A_1 E_2^\ast
- (B_2+E_2-E_3) D_1^\ast \Big] \nnb \\
\ek m_{\Lambda_b}
 \sqrt{\hat{r}_\Lambda} \hat{s}
\Big[ A_1 D_2^\ast + (A_2 - D_2 - D_3) D_1^\ast - B_1 E_2^\ast -
(B_2 + E_2 + E_3) E_1^\ast \Big] \nnb \\
\ek m_{\Lambda_b}^2 
\hat{s} (1-\hat{r}_\Lambda)
(A_2 D_2^\ast - B_2 E_2^\ast)
- m_{\Lambda_b}^2                       
\hat{s}^2 (D_2 D_3^\ast + E_2 E_3^\ast )
\Bigg\}~, \\ \nnb \\ \nnb \\
\label{e7721}
P_{NT} \es \frac{64 m_{\Lambda_b}^4 \lambda v}{3 \Delta}
\mbox{\rm Im} \Bigg\{
(A_1 D_1^\ast +B_1 E_1^\ast)
+ m_{\Lambda_b}^2 \hat{s}
(A_2^\ast D_2 + B_2^\ast E_2)
\Bigg\}~, \\ \nnb \\ \nnb \\
\label{e7722}
P_{TN} \es - \frac{64 m_{\Lambda_b}^4 \lambda v}{3 \Delta}
\mbox{\rm Im} \Bigg\{
(A_1 D_1^\ast +B_1 E_1^\ast)
+ m_{\Lambda_b}^2 \hat{s}
(A_2^\ast D_2 + B_2^\ast E_2)
\Bigg\}~, \\ \nnb \\ \nnb \\
\label{e7723}
P_{NN} \es \frac{32 m_{\Lambda_b}^4}{3 \hat{s} \Delta}
\mbox{\rm Re} \Bigg\{
24 \hat{m}_\ell^2 \sqrt{\hat{r}_\Lambda} \hat{s}
( A_1 B_1^\ast + D_1 E_1^\ast ) \nnb \\
\ek 12 m_{\Lambda_b} \hat{m}_\ell^2 \sqrt{\hat{r}_\Lambda} \hat{s}  
(1-\hat{r}_\Lambda +\hat{s}) (A_1 A_2^\ast + B_1 B_2^\ast) \nnb \\
\ar 6 m_{\Lambda_b} \hat{m}_\ell^2 \hat{s} \Big[ 
m_{\Lambda_b} \hat{s} (1+\hat{r}_\Lambda-\hat{s})
\Big(\vel D_3 \ver^2 + \vel E_3 \ver^2 \Big) +
2 \sqrt{\hat{r}_\Lambda} (1-\hat{r}_\Lambda+\hat{s})
(D_1 D_3^\ast + E_1 E_3^\ast)\Big] \nnb \\        
\ar 12 m_{\Lambda_b} \hat{m}_\ell^2 \hat{s} (1-\hat{r}_\Lambda-\hat{s})
(A_1 B_2^\ast + A_2 B_1 ^\ast + D_1 E_3^\ast + D_3 E_1^\ast) \nnb \\
\ek [ \lambda \hat{s} + 
2 \hat{m}_\ell^2 (1 + \hat{r}_\Lambda^2 - 2 \hat{r}_\Lambda + 
\hat{r}_\Lambda \hat{s} + \hat{s} - 2 \hat{s}^2) ]
\Big( \vel A_1 \ver^2 + \vel B_1 \ver^2 - \vel D_1 \ver^2 - 
\vel E_1 \ver^2 \Big) \nnb \\
\ar 24 m_{\Lambda_b}^2 \hat{m}_\ell^2 \sqrt{\hat{r}_\Lambda} \hat{s}^2 
(A_2 B_2^\ast + D_3 E_3^\ast)
- m_{\Lambda_b}^2 \lambda \hat{s}^2 v^2 
\Big( \vel D_2 \ver^2 + \vel E_2 \ver^2 \Big) \nnb \\
\ar m_{\Lambda_b}^2 \hat{s} \{ \lambda \hat{s} -
2 \hat{m}_\ell^2 [2 (1+ \hat{r}_\Lambda^2) - \hat{s} (1+\hat{s})
- \hat{r}_\Lambda (4+\hat{s})]\}
\Big( \vel A_2 \ver^2 + \vel B_2 \ver^2 \Big)
\Bigg\}~, \\ \nnb \\ \nnb \\
\label{e7724}
P_{TT} \es \frac{32 m_{\Lambda_b}^4}{3 \hat{s} \Delta}
\mbox{\rm Re} \Bigg\{
- 24 \hat{m}_\ell^2 \sqrt{\hat{r}_\Lambda} \hat{s}
( A_1 B_1^\ast + D_1 E_1^\ast ) \nnb \\
\ek 12 m_{\Lambda_b} \hat{m}_\ell^2 \sqrt{\hat{r}_\Lambda} \hat{s}  
(1-\hat{r}_\Lambda +\hat{s}) (D_1 D_3^\ast + E_1 E_3^\ast)
- 24 m_{\Lambda_b}^2 \hat{m}_\ell^2 \sqrt{\hat{r}_\Lambda} \hat{s}^2
( A_2 B_2^\ast + D_3 E_3^\ast ) \nnb \\
\ek 6 m_{\Lambda_b} \hat{m}_\ell^2 \hat{s} \Big[ 
m_{\Lambda_b} \hat{s} (1+\hat{r}_\Lambda-\hat{s})
\Big(\vel D_3 \ver^2 + \vel E_3 \ver^2 \Big) -
2 \sqrt{\hat{r}_\Lambda} (1-\hat{r}_\Lambda+\hat{s})
(A_1 A_2^\ast + B_1 B_2^\ast)\Big] \nnb \\
\ek 12 m_{\Lambda_b} \hat{m}_\ell^2 \hat{s} (1-\hat{r}_\Lambda-\hat{s})
(A_1 B_2^\ast + A_2 B_1 ^\ast + D_1 E_3^\ast + D_3 E_1^\ast) \nnb \\
\ek [ \lambda \hat{s} - 
2 \hat{m}_\ell^2 (1 + \hat{r}_\Lambda^2 - 2 \hat{r}_\Lambda + 
\hat{r}_\Lambda \hat{s} + \hat{s} - 2 \hat{s}^2) ]
\Big( \vel A_1 \ver^2 + \vel B_1 \ver^2 \Big) \nnb \\
\ar m_{\Lambda_b}^2 \hat{s} \{ \lambda \hat{s} +
\hat{m}_\ell^2 [4 (1- \hat{r}_\Lambda)^2 - 2 \hat{s} (1+\hat{r}_\Lambda)         
- 2 \hat{s}^2 ]\}
\Big( \vel A_2 \ver^2 + \vel B_2 \ver^2 \Big) \nnb \\
\ar \{ \lambda \hat{s} -
2 \hat{m}_\ell^2 [5 (1- \hat{r}_\Lambda)^2 - 7 \hat{s} (1+\hat{r}_\Lambda)         
+ 2 \hat{s}^2 ]\}                                                               
\Big( \vel D_1 \ver^2 + \vel E_1 \ver^2 \Big) \nnb \\
\ek m_{\Lambda_b}^2 \lambda \hat{s}^2 v^2 
\Big( \vel D_2 \ver^2 + \vel E_2 \ver^2 \Big)
\Bigg\}~.
\eea
Explicit expression of $\Delta$ appearing in $P_{ij}$ can be found in
\cite{R7724}.

\section{Numerical results}

In this section we present our numerical results for the 
double--polarization asymmetries. 
The values of the input parameters we need in performing the numerical
calculations are: $\vel V_{tb} V_{ts}^\ast \ver = 0.0385$, 
$m_\tau = 1.77~GeV$, $m_\mu = 0.106~GeV$,
$m_b = 4.8~GeV$ \cite{R7725}, $m_t=172.7~GeV$ \cite{R7726} and $\tau_{B_0} =
(1.527 \pm 0.008)~ps$.

The $\Lambda_b \rar \Lambda$ transition form factors are the main input
parameters in performing the numerical analysis, which are embedded into the
expressions of the double--lepton polarization asymmetries. The analysis of
all form factors responsible for the $\Lambda_b \rar \Lambda$ transition has
not been accomplished so far. Therefore, for the form factors  we will use the 
results coming from QCD sum rules in corporation with HQET
\cite{R7727,R7728}, which reduce the number of independent form factors to
two, and their $q^2$ dependence are given in terms of three--parameter fit
as follows:
\bea
F_i(\hat{s}) = \frac{F(0)}{\ds 1-a_F \hat{s} + b_F \hat{s}^2}~. \nnb
\eea
The values of the parameters $F(0),~a_F$ and $b_F$ are given in table 1.
\begin{table}[h]    
\renewcommand{\arraystretch}{1.5}
\addtolength{\arraycolsep}{3pt}
$$
\begin{array}{|l|ccc|}  
\hline
& F(0) & a_F & b_F \\ \hline
F_1 &
\phantom{-}0.462 & -0.0182 & -0.000176 \\
F_2 &
-0.077 & -0.0685 &\phantom{-}0.00146 \\ \hline
\end{array}
$$
\caption{Form factors for $\Lambda_b \rar \Lambda \ell^+ \ell^-$
decay in a three parameter fit.}
\renewcommand{\arraystretch}{1}
\addtolength{\arraycolsep}{-3pt}
\end{table}  

The analysis of the double--lepton polarization asymmetries leads to the
following results:

\begin{itemize}

\item $P_{LL}$ in UED for the $\Lambda_b \rar \Lambda \mu^+ \mu^-$ decay 
practically coincides with the SM result, for all values of $q^2$.

\item For the $\Lambda_b \rar \Lambda \tau^+ \tau^-$ case the difference
between the predictions of SM and UED is substantial at $q^2 = 12~GeV^2$,
i.e., $(P_{LL})_{UED} = 2 (P_{LL})_{SM}$ at $1/R=200~GeV$; with increasing
$q^2$ the difference between the two models decreases, but they never
coincide (see Fig. 1).   

\item For the $\Lambda_b \rar \Lambda \mu^+ \mu^-$ decay, 
starting from $q^2 = 1~GeV^2$ up to the end of the spectrum, the value of
$P_{TT}$ in the SM is larger compared to that of the one predicted by the UED 
model. Especially, up to $q^2 = 10~GeV^2$, 
$(P_{TT})_{SM} = 2 (P_{TT})_{UED}$ (see Fig. 2).

Therefore measurement of the values of $P_{LL}$ for the $\Lambda_b \rar
\Lambda \tau^+ \tau^-$ decay and $P_{TT}$ for the $\Lambda_b \rar
\Lambda \mu^+ \mu^-$ decay can give quite important information about the
presence of new physics beyond the SM.
 
\item  For the $\Lambda_b \rar \Lambda \tau^+ \tau^-$ decay, the
difference between the predictions of the SM and UED is maximally about
$60\%$, i.e., in terms of modulo, $\vel (P_{LT})_{UED} \ver > \vel
(P_{LT})_{SM} \ver$, which can also be very useful for establishing new
physics (see Fig. 3).

\item The maximum value of the difference between the SM and UED models
concerning $P_{TN}$, $P_{NT}$, $P_{LN}$, $P_{NL}$, $P_{TL}$  
(excluding $q^2 = 1~GeV^2$ region for the $\Lambda_b \rar \Lambda \mu^+
\mu^-$ channel) for both decay
channels, is about $10\%$. Note that at $q^2 = 1~GeV^2$, $(P_{LT})_{UED} =
(P_{TL})_{UED} \simeq 2 (P_{LT})_{SM} = 2 (P_{TL})_{SM}$, for the $\Lambda_b
\rar \Lambda \mu^+ \mu^-$ decay.

\item When $2~GeV^2 \le q^2 \le 10~GeV^2$, the prediction of the UED model
on $P_{NN}$ is maximally two times larger than the SM prediction for the
$\Lambda_b \rar \Lambda \mu^+ \mu^-$ decay (see Fig. 4).      

\item As far as $P_{NN}$ for the $\Lambda_b \rar \Lambda \tau^+ \tau^-$ decay
is concerned, the situation is more promising. When the momentum transfer
square $q^2$ varies in the region $14~GeV^2 \le q^2 \le 18~GeV^2$, the
difference between the results of the two models on $P_{NN}$ is quite
large, about four times, i.e., $(P_{NN})_{UED} \simeq 4 (P_{NN})_{SM}$,
and the magnitude of $P_{NN}$ is larger more than $10\%$ in the UED model
compared to that in the SM, which can be measurable in the experiments.
  
\end{itemize}
   
From the above--presented discussion we conclude that measurement of various
double--lepton polarization asymmetries can be very useful for establishing
new physics predicted by the UED model. Here we should note that
single--lepton polarization is not a suitable tool for discrimination of the
UED model and SM (see \cite{R7719}).

In conclusion, we study the double--lepton polarization asymmetries
in the UED model. We find that various double--lepton polarization
asymmetries are very sensitive to the UED model and the results are
substantially different compared to the ones obtained in the SM, and hence
can serve as a promising tool for establishing new physics beyond the SM.

\section*{Acknowledgments}

One of the authors (T. M. A) is grateful to T\"{U}B\.{I}TAK for partially
support of this work under the project 105T131.

\newpage

\section*{Figure captions}
{\bf Fig. 1} The dependence of $P_{LL}$ on $q^2$ for the $\Lambda_b \rar
\Lambda \tau^+ \tau^-$ decay at fixed values of the compactification
parameter $1/R$. For completeness, here and in all 
following figures, SM results are also given. \\ \\
{\bf Fig. 2} The dependence of $P_{TT}$ on $q^2$ for the $\Lambda_b \rar
\Lambda \mu^+ \mu^-$ decay at fixed values of the compactification
parameter $1/R$. \\ \\
{\bf Fig. 3} The same as in Fig. 1, but for $P_{LT}$. \\ \\
{\bf Fig. 4} The same as in Fig. 2, but for $P_{NN}$. \\ \\
{\bf Fig. 4} The same as in Fig. 1, but for $P_{NN}$. \\ \\

\newpage

\begin{figure}
\vskip 3. cm
    \includegraphics{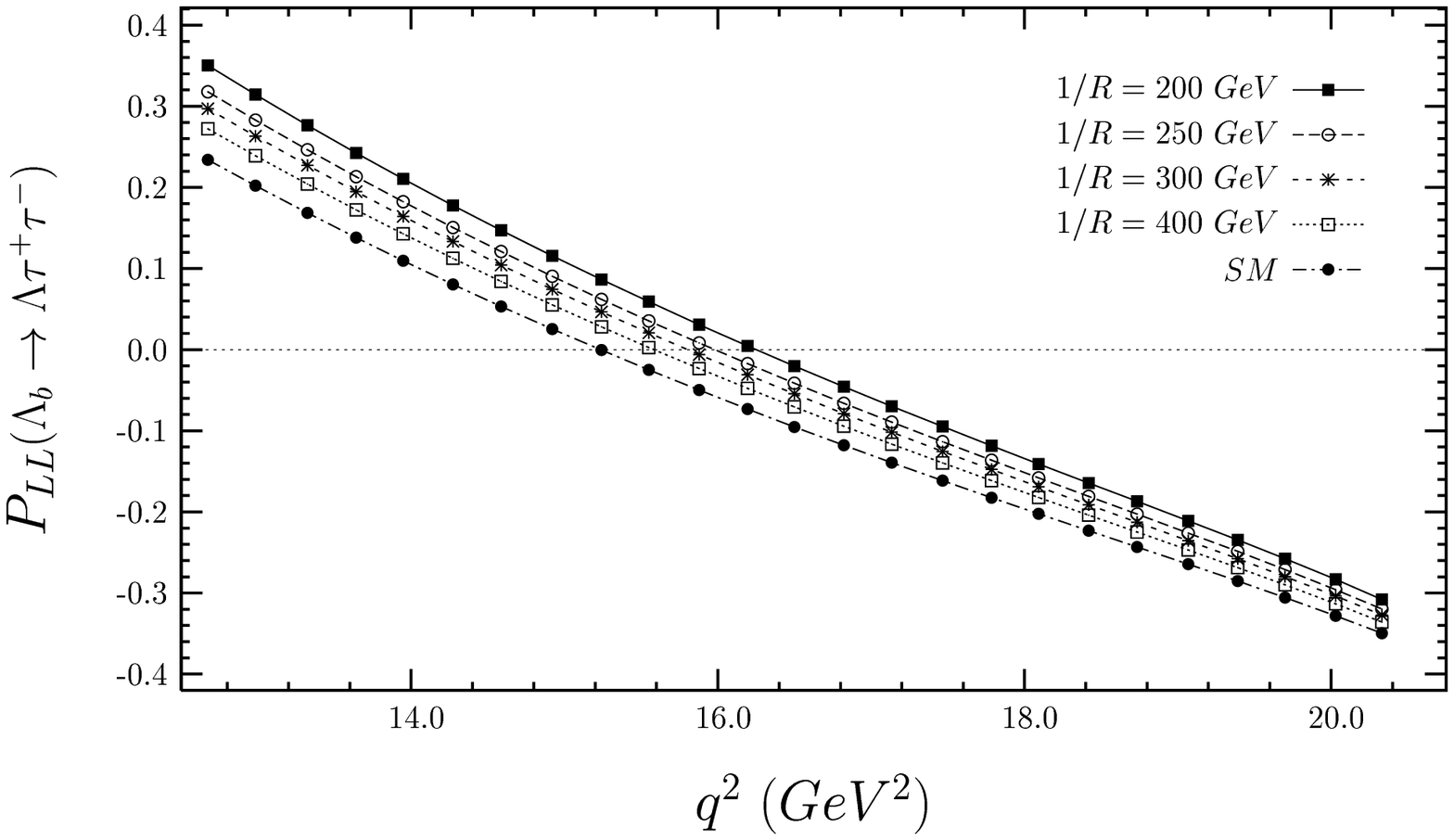}
\vskip 6.3cm
\caption{}
\end{figure}

\begin{figure}
\vskip 4.0 cm
    \includegraphics{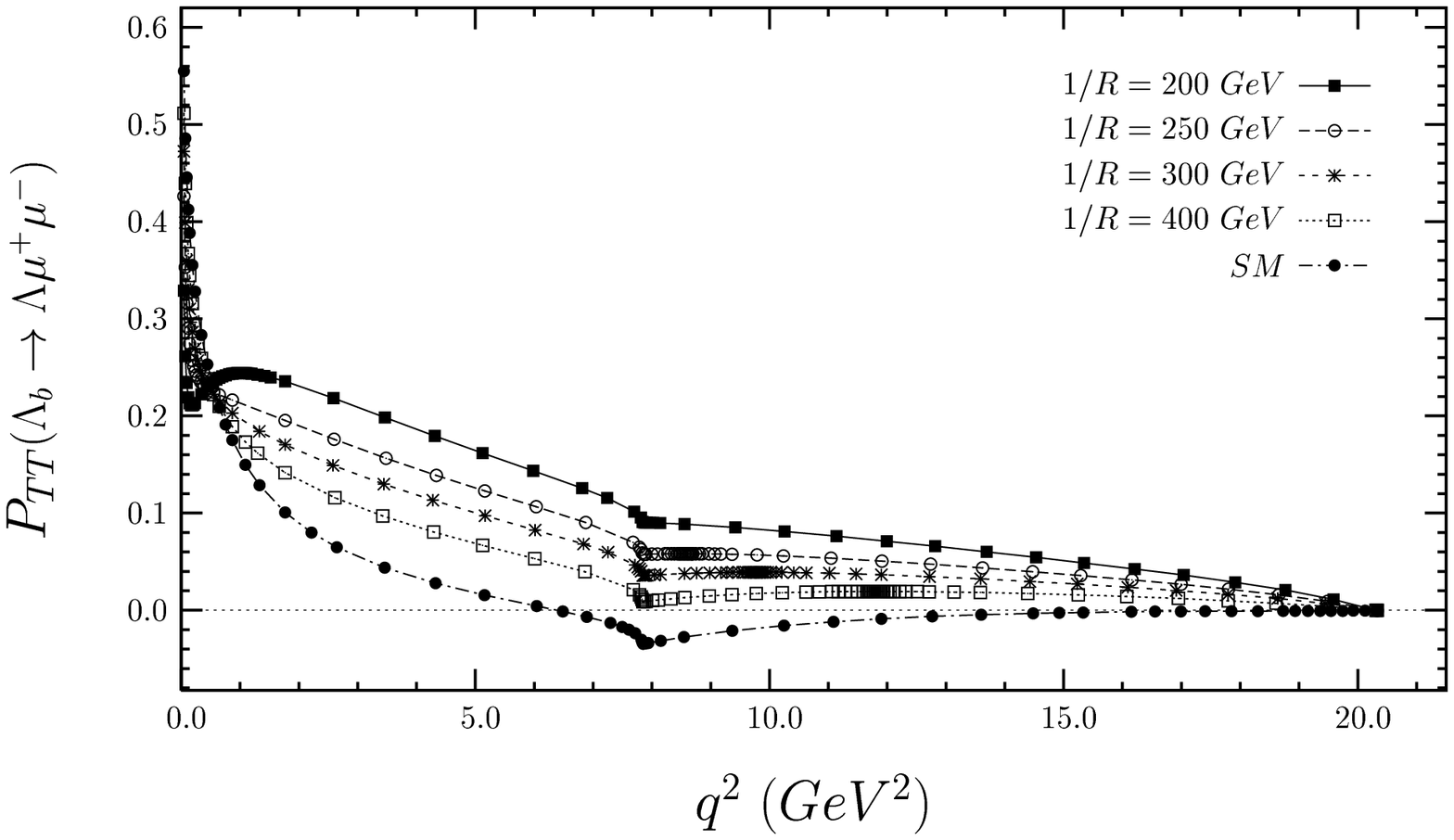}
\vskip 6.3 cm
\caption{}
\end{figure}

\begin{figure}
\vskip 3. cm
    \includegraphics{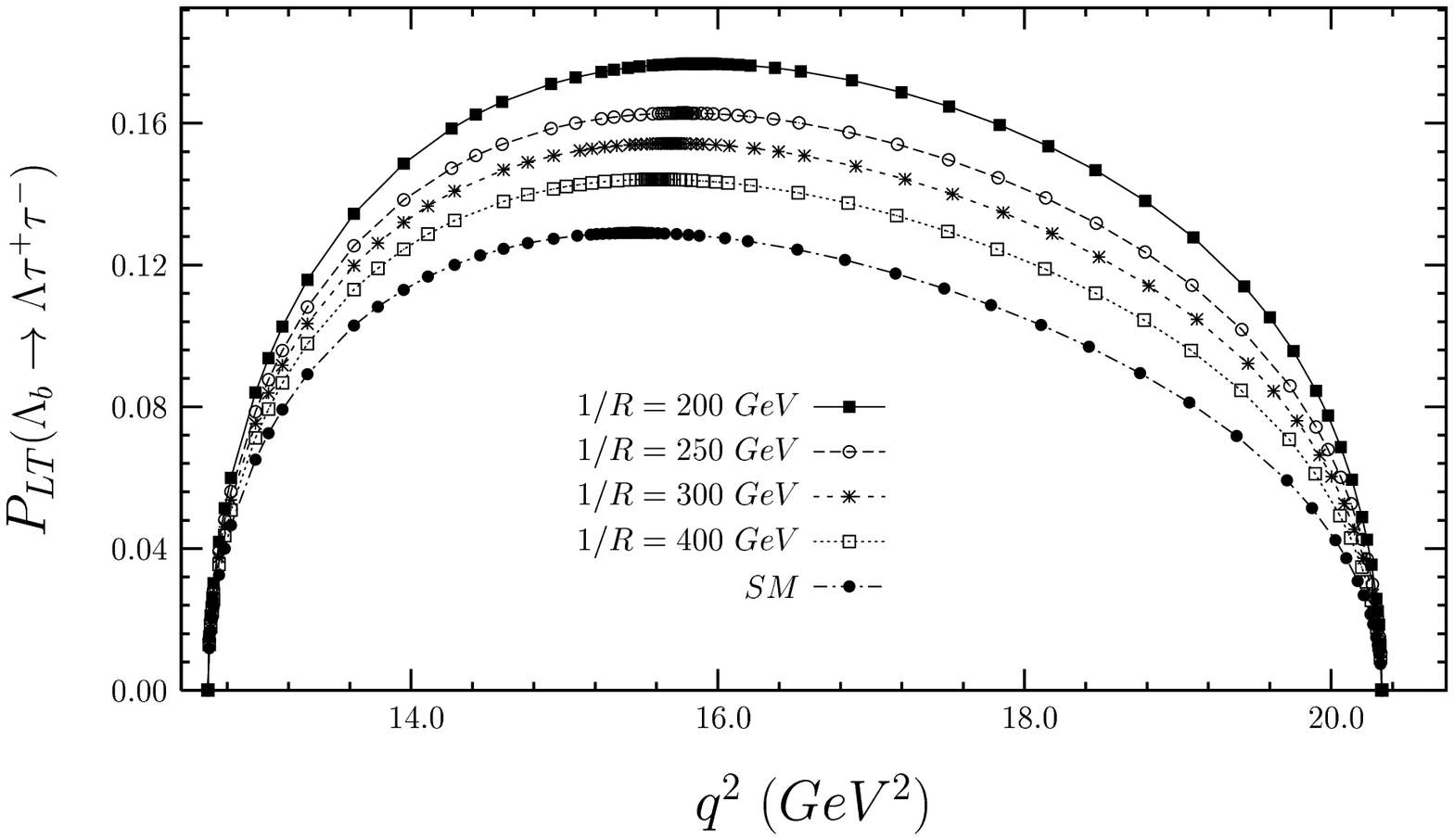}
\vskip 6.3cm
\caption{}
\end{figure}

\begin{figure}
\vskip 4.0 cm
    \includegraphics{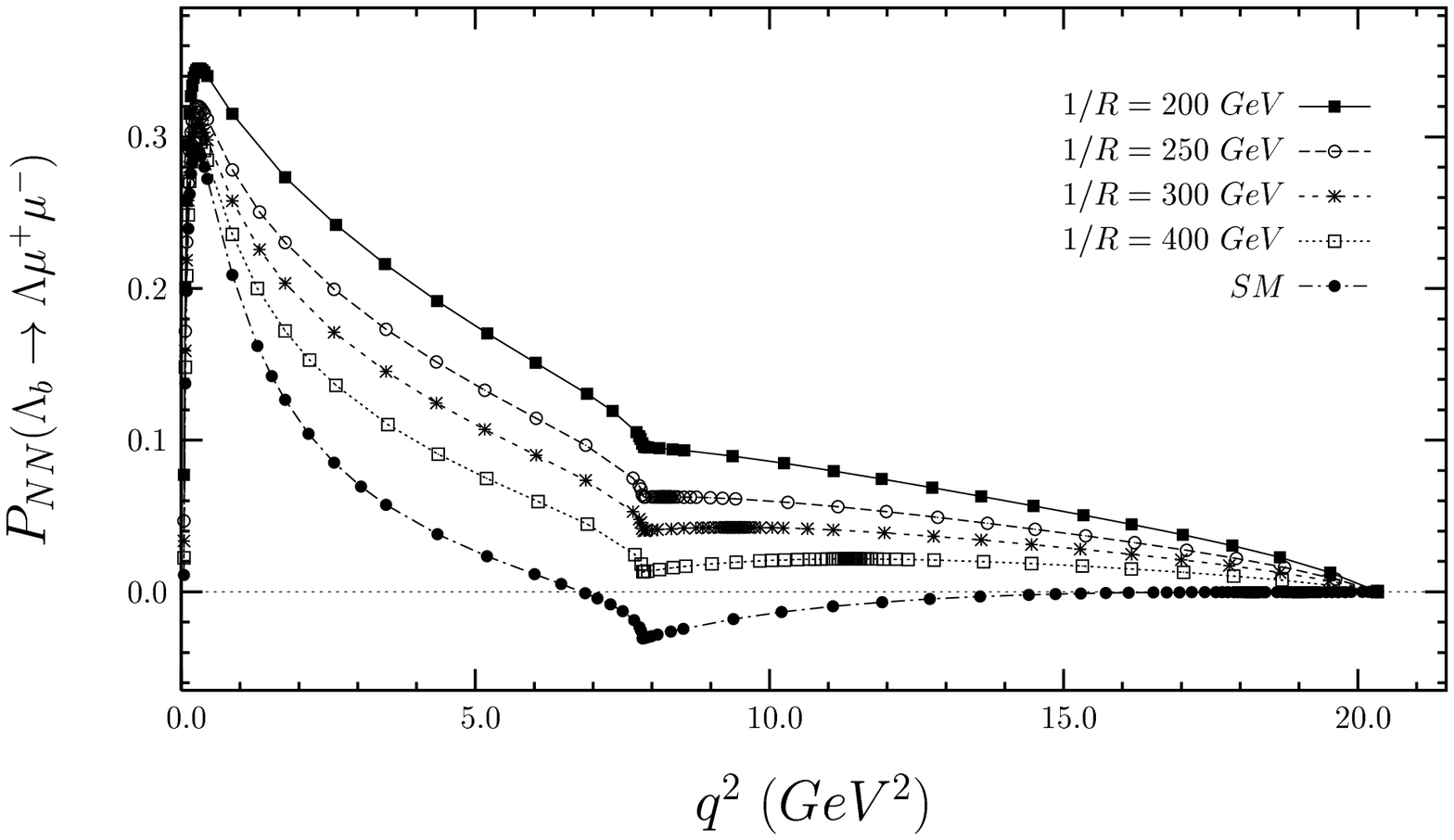}
\vskip 6.3 cm
\caption{}
\end{figure}

\begin{figure}
\vskip 3. cm
    \includegraphics{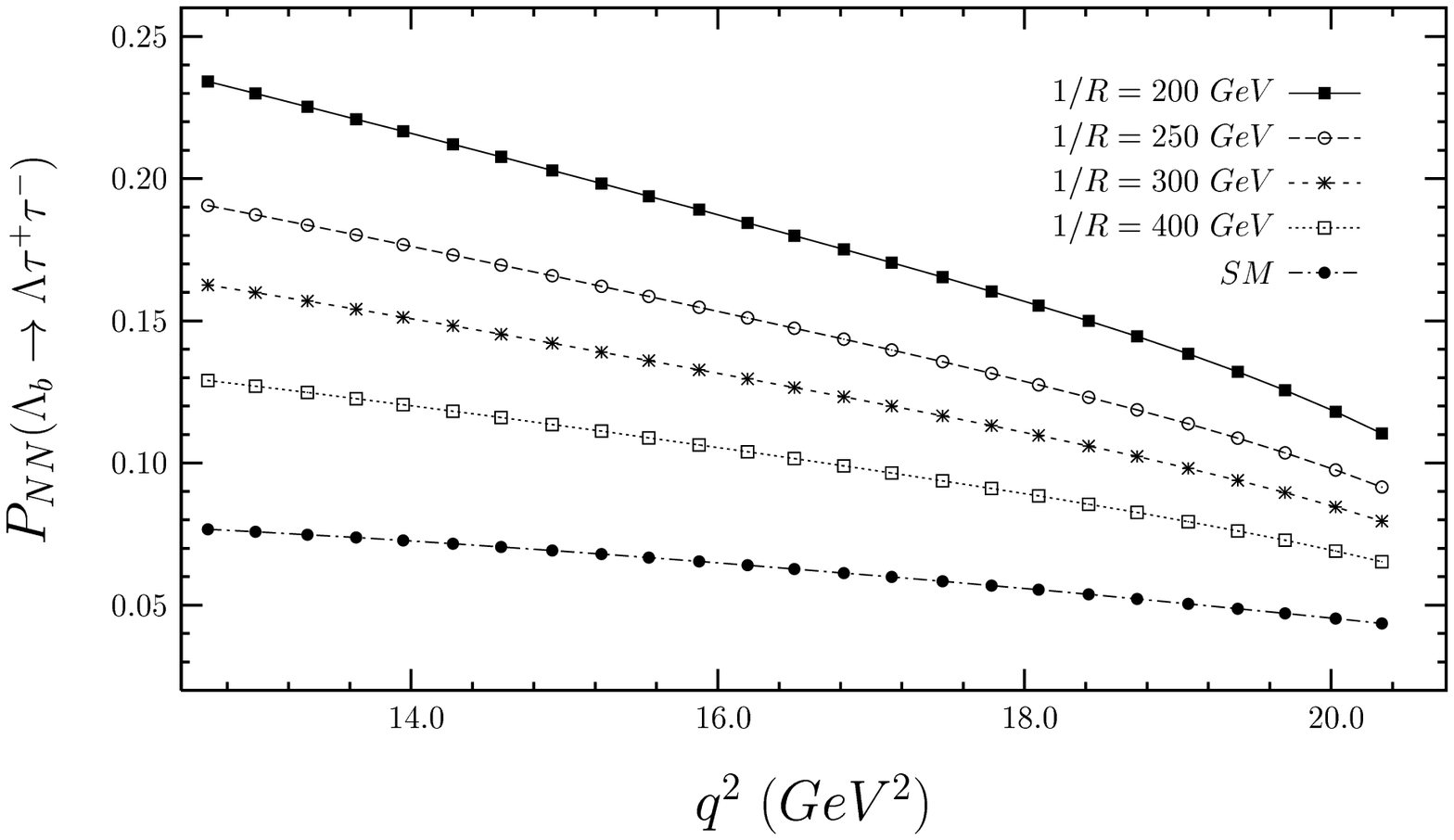}
\vskip 6.3cm
\caption{}
\end{figure}

\end{document}